\documentclass[preprint,superscriptaddress,nofootinbib]{revtex4}

\usepackage{graphicx}
\usepackage{amssymb}

\begin{document}

\def\be{\begin{equation}}
\def\ee{\end{equation}}
\def\bea{\begin{eqnarray}}
\def\eea{\end{eqnarray}}
\def\met{\slash{\!\!\!\!E}_T}
\preprint{ ANL-HEP-PR-11-15, EFI-11-7, IPMU11-0030}

\title{Top Quark Polarization As A Probe of Models \\with Extra Gauge Bosons}

\author{Edmond L. Berger}
\email{berger@anl.gov}
\affiliation{High Energy Physics Division, Argonne National Laboratory, Argonne, IL 60439, U.S.A}

\author{Qing-Hong Cao}
\email{caoq@hep.anl.gov}
\affiliation{High Energy Physics Division, Argonne National Laboratory, Argonne, IL 60439, U.S.A}
\affiliation{Enrico Fermi Institute, University of Chicago, Chicago, Illinois 60637, U.S.A.}

\author{Chuan-Ren Chen}
\email{chuan-ren.chen@ipmu.jp}
\affiliation{Institute for Physics and Mathematics of the Universe, University of Tokyo, Chiba 277-8568, Japan}

\author{Hao Zhang}
\email{haozhang.pku@pku.edu.cn}
\affiliation{Department of Physics and State Key Laboratory of Nuclear Physics and Technology, Peking University, Beijing 100871, China}

\begin{abstract}
New heavy gauge bosons exist in many models of new physics beyond the standard model 
of particle physics.   Discovery of  these $W^{\prime}$ and $Z^{\prime}$ resonances 
and the establishment of their spins, couplings, and other quantum numbers would shed 
light on the gauge structure of the new physics.  The measurement of the polarization of the 
SM fermions from the gauge boson decays would decipher the handedness of the coupling 
of the new states, an important relic of the primordial new physics symmetry.    Since the top 
quark decays promptly, its decay preserves spin information.  We show how decays of new 
gauge bosons into third generation fermions ($W^\prime \to tb,~Z^\prime\to t\bar{t}$) can be used to determine the handedness of 
the couplings of the new states and to discriminate among various new physics models. 
\end{abstract}

\maketitle

\section{Introduction}

One of the missions of the Large Hadron Collider (LHC) is to piece together
some of nature's original symmetries.  The search for these symmetries is a part of
the ultimate quest to unify all of the particles and forces within a grand unified theory 
that exhibits overarching gauge symmetries.  Theoretical clues to the original state 
of symmetry may be present in conserved or nearly conserved quantities in nature 
today.   As remnants of symmetry breaking, extra gauge bosons exist in many models 
of new physics (NP) that go beyond the standard model (SM).   The discovery of new 
neutral and charged gauge bosons and the establishment of their quantum numbers 
would shed light on the gauge structure of NP~\cite{Robinett:1981yz,
Langacker:1984dc,Barger:1986hd,London:1986dk,Rosner:1986cv,Marciano:1990dp,
Rosner:1995ft,Rosner:1996eb, Carena:2004xs,
Rizzo:2007xs,Petriello:2008zr,Langacker:2008yv,Chiang:2009kb,delAguila:2009gz,
Diener:2010sy,Frank:2010cj,Grojean:2011vu}.
 
One salient property of new gauge bosons is the handedness of their couplings to 
SM fermions, whether dominantly left-handed as the SM $W$ and $Z$ vector bosons 
or possibly with large right-handed couplings.  In this paper we focus on new 
color-singlet $W^{\prime}$ and $Z^{\prime}$ production at the LHC and their decays 
into the third generation SM fermions $t,b$. We explore quantitatively the measurement 
of the chirality of the couplings of the new gauge bosons from the polarization of the top 
quarks in their decays.%
\footnote{The top quark polarization can also be used to probe new gauge bosons  
and scalars in exotic color representation such as sextet and anti-triplet; 
see  Ref.~\cite{Berger:2010fy,Zhang:2010kr} for details.}
The top quark is the only ``bare'' quark whose spin information 
can be measured from its decay products since the decay proceeds promptly via the 
weak interaction.  Among the top quark decay products, the 
charged lepton from $t \rightarrow b \ell \nu$ is the best analyzer of the top quark spin.  
In the helicity basis, the polarization of the top quark can be determined from the distribution 
in $\theta_\ell$, the angle of the lepton in the rest frame of top quark relative to 
the top quark direction of motion in the overall center-of-mass (cm) frame.  The 
angular correlation of the lepton $\ell^+$ is ${1\over 2}(1\pm\cos\theta_\ell)$,  
with the ($+$) choice for right-handed and ($-$) for left-handed top-quarks~\cite{Jezabek:1994zv,Mahlon:1995zn}.   

In addition to the matter of handedness of couplings, there are other reasons to search for 
the $Z^\prime/W^\prime$ in $t\bar{t}$ and $tb$ 
events.  One is that searches in the leptonic decay modes would fail in the so-called 
leptophobic models because the $Z^\prime$ and $W^\prime$ bosons in these models 
do not couple to leptons.   Searches in dijet invariant mass distributions are valuable but 
cannot determine whether a dijet resonance is a $Z^\prime$ or $W^\prime$ boson because 
the jet charge is not measurable.   In such cases, the third generation quarks are necessary 
for charge determinations of the heavy resonances; for example, the $Z^{\prime}$ bosons 
decay into $t\bar{t}$ and $b\bar{b}$ pairs and the $W^{\prime}$ bosons into $t\bar{b}$ 
and $\bar{t}b$ pairs. 

In Sec.~II, we describe models of new physics that contain extra gauge bosons and show how 
patterns of symmetry breaking are manifest in the handedness of the couplings of the new 
gauge bosons to SM fermions.  We illustrate a few of the NP models from the current literature.  
This section also includes a summary of  the existing constraints on masses and couplings of new 
gauge bosons.   In Sec. III, we present $W^{\prime}$ and $Z^{\prime}$ production cross sections 
at the LHC, both the inclusive rates and the rates of interest to us with all branching fractions 
included.  The collider signatures we study are $\ell^+ \met b\bar{b}$ for the $W^{\prime +}$ and the semileptonic decay of $t\bar{t}$, namely $\ell^\pm \met j j b \bar{b}$, for the $Z^\prime$.   
The missing energy $\met$ is carried off by a neutrino in the top quark decay.  
The dominant backgrounds are also computed and assessments are presented for 
the $W^{\prime}$ and $Z^{\prime}$ discovery potential.  
After we impose kinematic cuts and reconstruct the final states, we conclude that a $Z^\prime$ resonance with mass $1$ TeV could be seen above the SM $t\bar{t}$ background with a 
statistical significance more than $5$ standard deviations ($5 \sigma$) for $10~{\text {fb}}^{-1}$ integrated luminosity 
at $14$ TeV, provided its coupling $g_V \equiv \sqrt {g_L^2 + g_R^2}$ to the SM quarks is about 
$0.4$, consistent with bounds from Tevatron searches in the dijet final state.  The $W^\prime$ 
signal can be much larger than the SM background if a coupling strength $0.4$ to the 
SM quarks is assumed.   For purposes of comparison, the couplings of the SM $W$ boson to SM quarks are
$g_L^{Wu\bar{d}}=0.461$ and those of the SM $Z$ boson are 
$g_L^{Zu\bar{u}}=-0.257$, $g_R^{Zu\bar{u}}=0.115$, $g_L^{Zd\bar{d}}=0.314$ and $g_R^{Zd\bar{d}}=-0.057$.

Section~IV is devoted to the measurement of the top quark polarization and the determination of 
the handedness of the new gauge bosons.  We apply our approach to three benchmark models, 
the sequential SM-like $W^\prime/Z^\prime$ model (SSM), the top-flavor model, and the left-right symmetric model (LRM). 
These models provide different predictions for the left-handed fraction of the coupling strengths of 
the new gauge bosons.  We show that  the coupling of a $W^\prime$ to $tb$ can be determined precisely, 
whereas the uncertainty is relatively large for a $Z^\prime$ to $t\bar{t}$, owing mainly 
to better statistics and smaller SM backgrounds in the $W^\prime$ case.  
For $m_{W^\prime} ~(\simeq m_{Z^\prime}) \sim 1$ TeV, our determinations of the handedness of the 
$W^\prime$ and $Z^\prime$ couplings allow the three benchmark models be separated to varying 
degrees with $100~{\text{fb}}^{-1}$ of accumulated data. 
With this large data sample, one can distinguish the $Z^\prime$ models if the central values of 
their top quark polarizations differ by $\sim 20\%$. 
For a leptophobic $Z^\prime$, one can differentiate among different models if the 
difference of the handedness of the coupling to SM quarks is $\gtrsim 10\%$ (for coupling 
strength $\sim 0.4$)). 
Our overall summary is found in Sec.~V.

\section{Models with extra gauge bosons}
\label{models}

\begin{figure}
\begin{center}
\includegraphics[scale=0.5,clip]{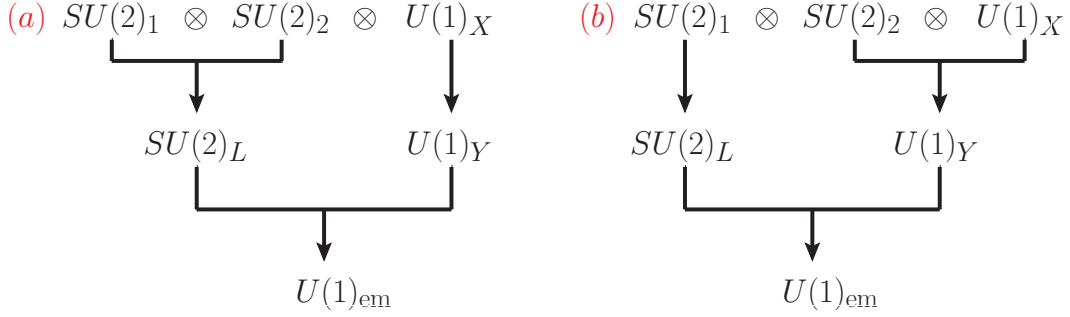}
\caption{Pictorial illustration of symmetry breaking patterns of $G(221)$ model.
\label{fig:221}}
\end{center}
\end{figure}

Extra gauge bosons may be classified according to their 
electromagnetic charges: $W^\prime$ (charged bosons) and $Z^\prime$ (neutral bosons).
While a $Z^\prime$ can originate from an additional abelian $U(1)$ group, 
a $W^\prime$ arises often in models with an extra non-abelian group.  
In this section we consider the so-called $G(221)$ model~\cite{Hsieh:2010zr} which carries the 
simplest non-abelian extension to the SM
\be
G(221) = SU(2)_1 \otimes SU(2)_2 \otimes U(1)_X.
\ee 
The model represents a typical gauge structure of many interesting NP models  
such as the non-universal model (NU)~\cite{Li:1981nk,Malkawi:1996fs,He:2002ha},
the ununified (UU) model~\cite{Georgi:1989ic, Georgi:1989xz}, 
the fermiophobic (FP) model~\cite{Barger:1980ix}, 
left-right (LR) models~\cite{Senjanovic:1975rk}, and so forth.  
Both $W^\prime$ and $Z^\prime$ bosons appear after the $G(221)$ symmetry is
broken to the SM symmetry $G_{SM}=SU(2)_L\otimes U(1)_Y$. 
As depicted in  Fig.~\ref{fig:221}, these models can be categorized by two symmetry breaking patterns,
\begin{itemize} 
\item[(a)] In the UU and NU models:\\
$U(1)_X$ is identified as the $U(1)_Y$ of the SM. The first stage of symmetry breaking 
$SU(2)_{1}\times SU(2)_{2}\to SU(2)_{L}$ occurs at the TeV scale, while the second stage of
symmetry breaking $SU(2)_{L}\times U(1)_Y \to U(1)_{em}$ occurs at the electroweak scale;
\item[(b)] In the FP and LR models: \\
$SU(1)_1$ is identified as the $SU(2)_L$ of the SM. The first stage of symmetry breaking 
$SU(2)_2 \times U(1)_{X}\to U(1)_{Y}$ occurs at the TeV scale, while the second stage of 
symmetry breaking $SU(2)_{L}\times U(1)_Y \to U(1)_{em}$ occurs at the electroweak scale.
\end{itemize}
In the first pattern the couplings of new gauge bosons to the SM fermions
are predominately left-handed while in the second pattern the couplings are 
right-handed.  A measurement of the polarization of the fermions
from the new gauge bosons would decipher the handedness of the couplings.

Rather than focusing on a specific model, we explore the discovery potential
of $W^\prime$ and $Z^\prime$ bosons in a model independent method, and we comment 
on a few new physics models later. 
The most general interaction of the $Z^\prime$ and $W^\prime$ to the SM quarks is 
\bea
\mathcal{L} &=& \bar{q}\gamma^{\mu}(g^{Z^\prime}_L P_L + g^{Z^\prime}_R P_R) q~Z^{\prime}_{\mu}  \nonumber \\ 
            &+& \bar{q}\gamma^{\mu}(g^{W^\prime}_L P_L + g^{W^\prime}_R P_R) q^\prime~W^{\prime+}_{\mu} + h.c.
            \label{eq:int}
\eea
where $P_{L/R}$ is the usual left- and right-handed projector and $q$ denotes the SM  quarks. 
The $Z^\prime$ and $W^\prime$ are understood here to be color singlet states, 
but one can easily obtain the interaction of color octet bosons, 
such as a $G^\prime$, from insertion of the $SU(3)_C$ color matrices $\lambda_A/2$
in Eq.~(\ref{eq:int}). 

The couplings $g_{L/R}^{Z^\prime}$ and $g_{L/R}^{W^\prime}$ usually are not independent
when the $W^\prime$ and $Z^\prime$ originate from the same gauge group.  
For example, in the left-right model, the SM right-handed quark singlets form a doublet $(u_R,d_R)$
which is gauged under the additional $SU(2)_R$ group. The $W^\prime$ and $Z^\prime$ transform as 
a $SU(2)_R$ triplet and their couplings to the SM quarks are correlated. 
In this work we first treat $g^{W^\prime}_{L/R}$ and $g^{Z^\prime}_{L/R}$ as independent in our 
collider simulation to derive the experimental sensitivity on $Z^\prime$ and $W^\prime$ measurements.
We then consider the correlation between the two couplings in the context of some NP models. 
We use $g_{L/R}$ to denote the left-handed and right-handed couplings of the $W^\prime$ and 
$Z^\prime$ to the SM quarks.  
For simplicity we assume the couplings of $Z^\prime$ to up- and down-type quarks are the same. 

For illustration we study three benchmark NP models in this work:
\begin{itemize}
\item sequential SM-like $W^\prime/Z^\prime$ (SSM) model: the $W^\prime$ and $Z^\prime$ 
couplings to SM fermions  
are exactly the same as the SM $W$ and $Z$ boson, and $m_{W^\prime}=m_{Z^\prime}$.  Although it is difficult in a realistic model to have couplings which are the same as in the SM, we show such a case for comparison. 
\item top-flavor model~\cite{Malkawi:1996fs}: the $W^\prime$ and $Z^\prime$ couplings are purely left-handed, and $m_{W^\prime}=m_{Z^\prime}$.
\item left-right symmetric model (LRM)~\cite{Senjanovic:1975rk}: Here, we consider a $SU(2)_L\times SU(2)_R\times U(1)_{B-L}$ model. The $W^\prime$ couplings to SM quarks are
purely right-handed while the $Z^\prime$ couplings are dominantly right-handed.

\end{itemize}
Table~\ref{table:coup} is a summary of the couplings of a $W^\prime$ and a $Z^\prime$ to SM third generation quarks in these models. The couplings to the quarks of first two generations are the same, except that one should replace $\sin\tilde{\phi}$ by $\cos\tilde{\phi}$ in the top-flavor model.   

\begin{table}
\caption{Couplings of a $W^\prime$ to $tb$ and a $Z^\prime$ to $t\bar{t}$ for the sequential SM-like $W^\prime/Z^\prime$ (SSM) model,  the left-right symmetric model (LRM), and the top-flavor model, where $s_w~(c_w, ~t_w) =\sin\theta_w~(\cos\theta_w,~\tan\theta_w)$, $\theta_w$ is the weak mixing angle, $g_2=e/{s_w}$ is the weak coupling, 
$\alpha_{LR} \simeq 1.6$, and $\sin\tilde\phi$ is taken to be $1/\sqrt{2}$.\label{table:coup}}  
\begin{tabular}{lll}
\hline 
      & $W^{\prime}tb$ & $Z^{\prime}t\bar{t}$\tabularnewline
\hline 
SSM & $ \displaystyle{\frac{g_2}{\sqrt{2}}}\bar{b}\gamma_\mu P_L t  W^{\prime\mu} $   
    & $ \displaystyle{\frac{g_2}{6c_w}}\bar{t}\gamma_\mu(  (-3+4s_w^2) P_L + 4s_w^2  P_R  ) t Z^{\prime\mu} $ 
\vspace*{3mm}
\tabularnewline
LRM & $ \displaystyle{\frac{g_2}{\sqrt{2}}} \bar{b}\gamma_\mu P_R t  W^{\prime\mu}$ 
    & $ \displaystyle{\frac{g_2t_w}{6}}\bar{t}\gamma_\mu( \frac{1}{\alpha_{LR}}  P_L +  (\frac{1}{\alpha_{LR}}-3\alpha_{LR}) P_R  ) t Z^{\prime\mu}$
\vspace*{3mm}
\tabularnewline
Top-Flavor & $\displaystyle{\frac{g_2\sin\tilde{\phi}}{\sqrt{2}}} \bar{b}\gamma_\mu P_L t W^{\prime\mu} $
           & $\displaystyle{\frac{g_2\sin\tilde{\phi}}{\sqrt{2}}}\bar{t}\gamma_\mu P_L  t  Z^{\prime\mu}$
\tabularnewline
\hline 
\end{tabular}
\end{table}%

\subsection{Bounds on masses and couplings}
The masses and couplings of $Z^\prime$ and $W^\prime$ bosons are bounded 
by various low energy measurements (mainly via the four-fermion
operators induced by exchanges of new heavy gauge bosons) such as the precision 
measurements at the $Z$-pole at LEP-I~\cite{Nakamura:2010zzi}, the $W$-boson mass~\cite{Nakamura:2010zzi},
the forward-backward asymmetry in $b\bar{b}$ production at 
LEP-II~\cite{Nakamura:2010zzi},
$\nu e$ scattering~\cite{Vilain:1996yf}, 
atomic parity violation~\cite{Vetter:1995vf,Wood:1997zq,Guena:2004sq}, 
Moller scattering~\cite{Anthony:2005pm}, and so forth.  
The bounds are severe when new gauge bosons couple to leptons directly, but they can 
be relaxed for a leptophobic model, as analyzed in Ref.~\cite{Hsieh:2010zr}.  

Tevatron data place a lower bound about $1.1~\rm{TeV}$ on the mass of a $W^\prime$~\cite{Aaltonen:2010jj} and about $1.07~\rm{TeV}$ for a $Z^\prime$~\cite{Aaltonen:2011gp}, based on the charged lepton plus missing energy ($\ell^\pm {\not \! \!E}_T$) and $\mu^+\mu^-$ final states, respectively, 
with the assumption that the couplings between the $W^\prime/Z^\prime$ and the SM fermions are 
the same as those in the SM.   Searches for the $W^\prime$ and $Z^\prime$ at the Tevatron in 
dijet events yield lower bounds on $m_{W^\prime}$ and $m_{Z^\prime}$, assuming SM couplings, of $840$ GeV and $740$ GeV, respectively~\cite{Aaltonen:2008dn}.
Recent CMS and ATLAS collaboration searches for a $Z^\prime$ from dilepton final states and a $W^\prime$ from lepton plus missing energy events place lower bounds $m_{Z^\prime} > 1.14$ TeV 
(CMS)~\cite{Collaboration:2011wq}, along with $m_{W^\prime} > 1.58$ TeV (CMS)~\cite{Collaboration:2011dx} and 
$m_{W^\prime} > 1.49$ TeV (ATLAS)~\cite{Collaboration:2011fe}.  The analyses assume the 
$Z^\prime$ and $W^\prime$ have sequential standard model couplings. 
CMS and ATLAS also present searches for a resonance in dijet events which also constrain the masses of a $W^\prime$ and a $Z^\prime$, but the lower bounds are looser than the Tevatron results~\cite{:2010bc,Khachatryan:2010jd}. 
If the couplings between the $W^\prime/Z^\prime$ and the SM particles are not small, one  can expect the discovery of the these heavy resonances sooner or later at the LHC. 
Negative searches for a $W^\prime/Z^\prime$ through the $t\bar{b}$ and $t\bar{t}$ final states at the Tevatron impose upper bounds on the production cross section times decay branching ratio ($\sigma_{W^\prime} Br(W^\prime\to t\bar{b}),~\sigma_{Z^\prime} Br(Z^\prime\to t\bar{t})$) for masses up to 
$950$ GeV and $900$ GeV for $W^\prime$ and $Z^\prime$, respectively~\cite{cdftt,cdftb}.

%
%
%
%
\section{LHC phenomenology}

We divide our discussion of $W^{\prime\pm}$ and $Z^\prime$ phenomenology into 
two parts.   In this section, we present our evaluation of the production cross sections and discuss 
the pertinent backgrounds with a view toward understanding the discovery potential of the two states.  Section~IV is then devoted to an examination of top quark polarization measurements as a means to learn more about the models that produce $W^{\prime}$ and $Z^\prime$ bosons.  

\begin{figure}[t]
\begin{center}
\includegraphics[scale=0.7,clip]{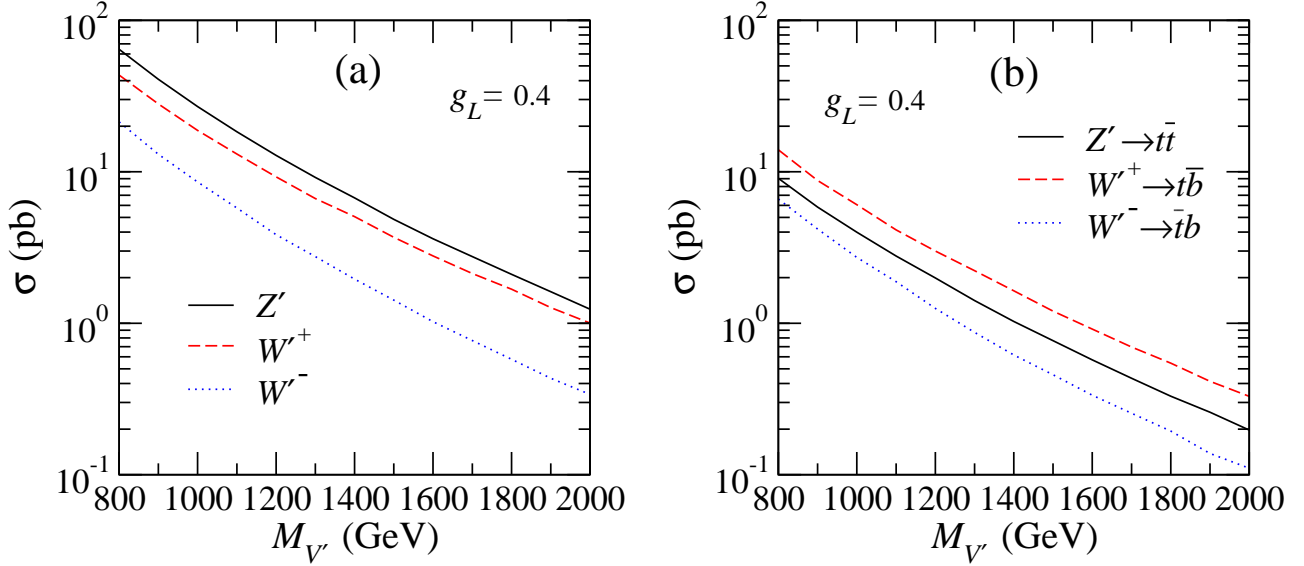}
\caption{Production cross sections of the $W^\prime$ and $Z^\prime$ at 14~TeV 
for the choice $g_L=0.4$ and $g_R=0$.   
The solid curves provide results before decay branching fractions are included, whereas the dashed 
curves include branching fractions.
\label{fig:lhc}}
\end{center}
\end{figure}

The $Z^\prime$ production cross section is 
\be
\sigma_{Z^\prime}(\hat{s})=\frac{\beta}{192\pi}
\frac{\hat{s}}{(\hat{s}-m_{Z^\prime}^2)^2+m^2_{Z^\prime}\Gamma^2_{Z^\prime}}
\left[(g_L^2 + g_R^2)^2(3+\beta^2)+ 6g_Lg_R(g_L^2+g_R^2)(1-\beta^2)\right],
\ee
where $\beta=\sqrt{1-4m_t^2/\hat{s}}$.
The $W^\prime$ production cross section is 
\be
\sigma_{W^\prime}(\hat{s})=\frac{(1-x_t^2)^2}{96\pi}
\frac{\hat{s}}{(\hat{s}-m_{W^\prime}^2)^2+m^2_{W^\prime}\Gamma^2_{W^\prime}}
(g_L^2 + g_R^2)^2\left(2+x_t^2\right),
\ee
where $x_t=m_t/\sqrt{\hat{s}}$.  
The partial decay width of $V^\prime \to q \bar{q}^\prime$ ($V^\prime=W^\prime,~Z^\prime$)  is 
\be
\Gamma(V^\prime \to q\bar{q}^\prime) =
\frac{m_{V^\prime}}{8\pi} \beta_0 
\left[(g_L^2 + g_R^2) \beta_1 + 6 g_L g_R \frac{m_q m_{q^\prime}}{m^2_{V^\prime}}\right],
\ee 
where
\be
\beta_0 = \sqrt{1 - 2\frac{m_q^2+m^2_{q^\prime}}{m^2_{V^\prime}}+
\frac{(m_q^2-m_{q^\prime}^2)^2}{m^4_{V^\prime}}},
\qquad
\beta_1 = 1 - \frac{m^2_q + m^2_{q^\prime}}{2m^2_{V^\prime}} 
-\frac{(m^2_q-m^2_{q^\prime})^2}{2m_{V^\prime}^4}.
\ee
Evaluations of the cross sections are presented  
in Fig.~\ref{fig:lhc} at the LHC with center of mass energy 14~TeV.  
In the mass range of interest to us the $W^\prime$ and $Z^\prime$ bosons are
much heavier than the top quark, and $m_t$ can be ignored. 
Hence, the decay branching ratio of $Z^\prime \to t\bar{t}$ is about 1/6 while 
the ratio of $W^\prime \to t \bar{b}$ is 1/3. 
The cross sections for other values of $g_L$ and $g_R$ can be obtained from
the curves in Fig.~\ref{fig:lhc} by a simple scaling, $\sigma_{V^\prime} \propto (g_L^2 + g_R^2)$.

We use MadGraph/MadEvent~\cite{Alwall:2007st} to obtain the signal and background distributions.
The widths of $W^\prime$ and $Z^\prime$ for different $g_L$ and $g_R$ couplings are calculated 
in BRIDGE~\cite{Meade:2007js}. These results are computed at leading order with 
the renormalization scale ($\mu_R$) and factorization scale ($\mu_F$) chosen as
\be
\mu_R = \mu_F = \sqrt{m_t^2 + 2 p_T^2(t)}.
\ee
The CTEQ6.1L parton distribution functions (PDFs)~\cite{Pumplin:2002vw} are used.

The coupling strength is set at 
$g_V \equiv \sqrt{g_L^2+g_R^2} = 0.4$, which respects the dijet constraints at the Tevatron~\cite{Aaltonen:2008dn} 
in the leptophobic $W^\prime/Z^\prime$ models with a universal coupling.
We plot the production cross sections with the choice of $g_L=0.4,~g_R=0$ in the figure.
The curves shown in Fig.~\ref{fig:lhc} also represent the cross sections for 
$g_R=0.4,~g_L=0$.

After including branching fractions, we also plot the cross 
sections for $pp\to Z^\prime X \to t\bar{t} X$,  $pp\to W^{\prime+} X\to t\bar{b} X$, and 
$pp\to W^{\prime-} X\to \bar{t}bX$.   Universal couplings of the $W^\prime/Z^\prime$ bosons to 
three generation of quarks are understood.   Because the $u$-quark parton density in the proton 
is large, it is easy to understand that the $Z^\prime$ has the largest production cross section while 
the $W^{\prime-}$ has the smallest one.

We explore the $W^\prime$ in the $t\bar{b}$ decay mode, and the $Z^\prime$ in the $t\bar{t}$ mode. 
To be able to measure the top quark polarization, we focus on final states in which the top quark 
decays leptonically.   Therefore, the $W^\prime$ search is via the channel 
$p p \to W^\prime X\to t b X\to \ell^\pm \nu b \bar{b} X$.  The $Z^\prime$ search is done in the 
channel $p p \to Z^\prime X\to t \bar{t} X\to \ell^\pm \nu j j b \bar{b} X$.   We consider semileptonic decay 
of only one of the top quarks in the $t\bar{t}$ mode of $Z^\prime$ decay because of its large branching ratio.   
One can also use the dilepton channel to search for a $Z^\prime$ using the MT2-assisted method 
discussed in Ref.~\cite{Berger:2010fy,Zhang:2010kr} to fully reconstruct the $Z^\prime$. 

\subsection{$Z^\prime$ discovery potential}
\label{sec:lhc_z}
The collider signature of interest to us for the $Z^\prime$ boson is $\ell^\pm\nu jj b\bar{b}$, where  one top quark decays semileptonically and the other decays hadronically.  We consider in this subsection only $\mu^+$  final states, $t \to b \mu^+ \nu_\mu$, but the statistics will increase when the different  flavors and charges of the leptons are combined.  The major SM background is $t\bar{t}$ production 
via the QCD interaction.  The sum of other backgrounds, such as $W/Z(\to \ell\ell)+$jets, single top, $W+b\bar{b}$, etc., is a factor of $\gtrsim 20$ smaller after the usual semileptonic $t\bar{t}$ selection 
cuts~\cite{Aad:2009wy}.  

At the analysis level, all signal and background events are required to pass the acceptance cuts listed here:
\bea
p_T(\ell,j) > 20~{\rm GeV}, &\quad |\eta(\ell, j)|<2.5,&\quad  \Delta R_{jj,j\ell} > 0.4,\nonumber \\
\met >30~{\rm GeV}, &\quad H_T > 500~{\rm GeV}, &\quad  M_T > 800~{\rm GeV}, 
\eea 
where $p_T$($\eta$, $\met$) denotes the transverse momentum (rapidity, missing transverse momentum), 
$\Delta R_{kl}\equiv\sqrt{\left(\eta_{k}-\eta_{l}\right)^{2}+\left(\phi_{k}-\phi_{l}\right)^{2}}$ 
is the separation in the azimuthal angle ($\phi$)-pseudorapidity 
($\eta$) plane between the objects $k$ and $l$,
$H_T$ is the scalar sum of the transverse momenta of the final state visible particles plus 
$\met$,  and $M_T$ is the cluster transverse 
mass defined as $M_T \equiv \sqrt{\sum_{i=j,\ell} p_i^2+\met^2} + \met $.
We model detector resolution effects by smearing the final state energy according to 
$\delta E/E= \mathcal{A}/\sqrt{E/{\rm GeV}}\oplus \mathcal{B}$,
where we take $\mathcal{A}=10(50)\%$ and $\mathcal{B}=0.7(3)\%$ for leptons (jets). To
account for $b$-jet tagging efficiencies, we demand two $b$-tagged jets, each with a tagging efficiency of $60\%$.
 
We consider two masses $m_{Z^\prime} =1$TeV and $1.5$TeV and the coupling strength
$g_{Z^\prime}=0.4$ to satisfy the bounds on heavy resonance searches in the dijet channel 
at the Tevatron.  The cross sections for $t\bar{t}$ before and after the cuts are shown in 
Table~\ref{table:xszprime}.
\begin{table}
\caption{Cross sections (in fb) for the signal process $pp\to Z^\prime \to t\bar{t} \to \mu^+ \met j j b \bar{b} $ and the SM backgrounds at 14~TeV. 
Two $b$-jets are tagged.  ``With cuts" means the cross sections after all of the kinematic cuts, b-tagging and reconstruction.  The universal coupling of the $Z^\prime$ to the SM quarks is set to be $0.4$. The mass window cuts are $\Delta M = 150~(200)$ GeV for a $1~(1.5)$~TeV $Z^\prime$ resonance, respectively.}
\begin{tabular}{c|ccc|ccc|ccc}
\hline
   &  & $Z_R^\prime$&  & &$Z_L^\prime$& & &Background & \tabularnewline
\hline  
$M_{Z^\prime}$& No cut & With cuts & $\Delta M$ & No cut & With cuts & $\Delta M$ & No cut             & With cuts & $\Delta M$ \tabularnewline
\hline
1~TeV         & 275.6   & 29.5      & 28.3       &  275.6  & 27.2 	   & 26.2       & $3.75\times 10^4$  & 133.1     & 87.0 \tabularnewline
1.5~TeV       & 51.4   & 3.0       & 2.6        & 52.5   & 3.9       & 3.5        & $3.75\times 10^4$  & 133.1     & 11.3 \tabularnewline
\hline
\end{tabular}
\label{table:xszprime}
\end{table}%

Since there is only one neutrino in the final state,  the undetected $z$ component of the neutrino momentum can be reconstructed from the on-shell condition of the $W$-boson.  This procedure 
leads to a two-fold solution, but the ambiguity can be removed by the top quark on-shell condition 
$m_{b\ell\nu}^2 = m_t^2$, where $m_t=173.3$ GeV~\cite{tevatron:1900yx} is used.  If no real solution is obtained, we use the top quark on-shell condition first, and then the $W$-boson on-shell condition 
to choose the better solution.   
Since the $b$ and $\bar{b}$ are indistinguishable, one must pick which $b$-jet should be 
combined with the charged lepton.  For this, we use the on-shell condition of the top quark 
in the hadronic decay to select one of the $b$-jets to pair with the two jets that are the 
decay products of $W$-boson.  The efficiency for choosing the correct $b$-jets  to reconstruct 
the semileptonic and hadronic decays of the top quarks can reach $\gtrsim99.7\%$. 

\begin{figure}[t]
\begin{center}
\includegraphics[scale=0.45,clip]{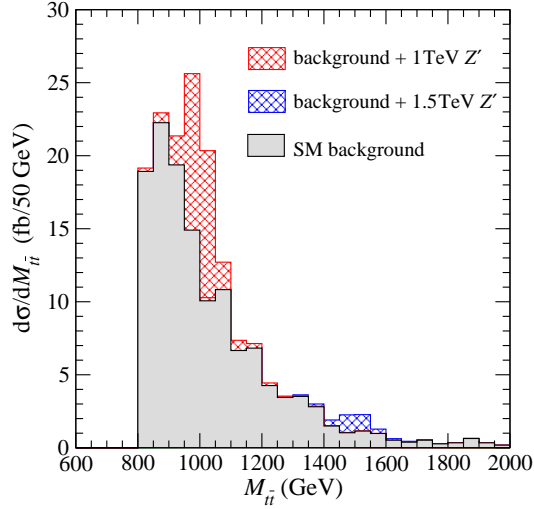}
\caption{Distribution in $m_{t\bar{t}}$ for a $Z^\prime$ and the SM $t\bar{t}$ background at the LHC. The coupling strength $g_{Z^\prime}$ is set to be $0.4$. }
\label{fig:mtt}
\end{center}
\end{figure}

In our simulations, after kinematic cuts and event reconstruction, the $Z^\prime$ resonance peak 
can be seen above the SM continuum in the $ t\bar{t}$ invariant mass distribution, $m_{t\bar{t}}$, 
especially for a 1 TeV $Z^\prime$ .  The results are shown in Fig.~\ref{fig:mtt}.   To enhance the discovery significance, we further focus on the events within a mass window around the resonance:
\bea
|m_{t\bar{t}} - m_{Z^\prime}| \leq 150~(200)~{\rm GeV,~ for~1~(1.5)~TeV~Z^{\prime}}
\eea 
\begin{figure}[t]
\begin{center}
\includegraphics[scale=0.4,clip]{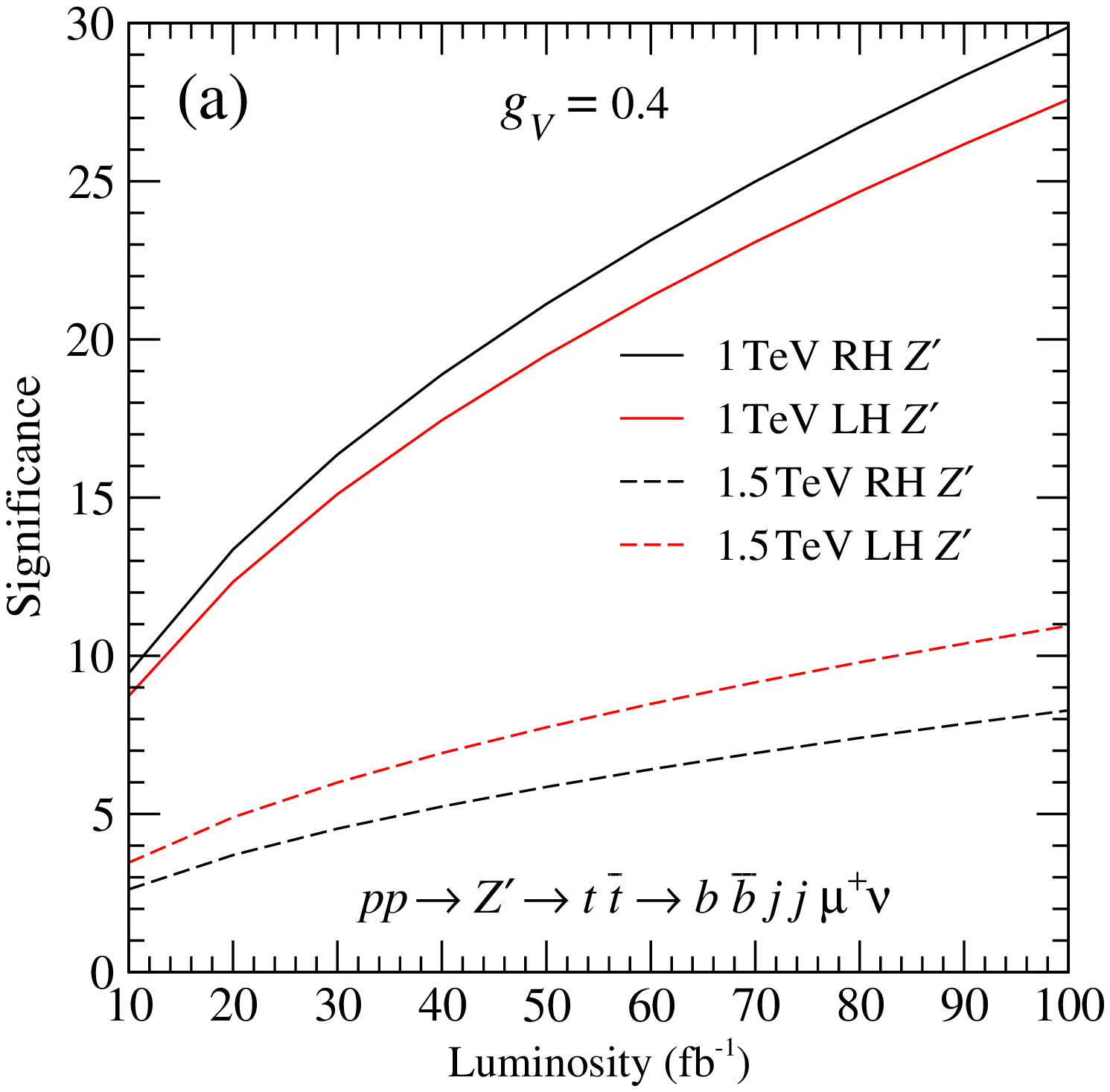}
\includegraphics[scale=0.4,clip]{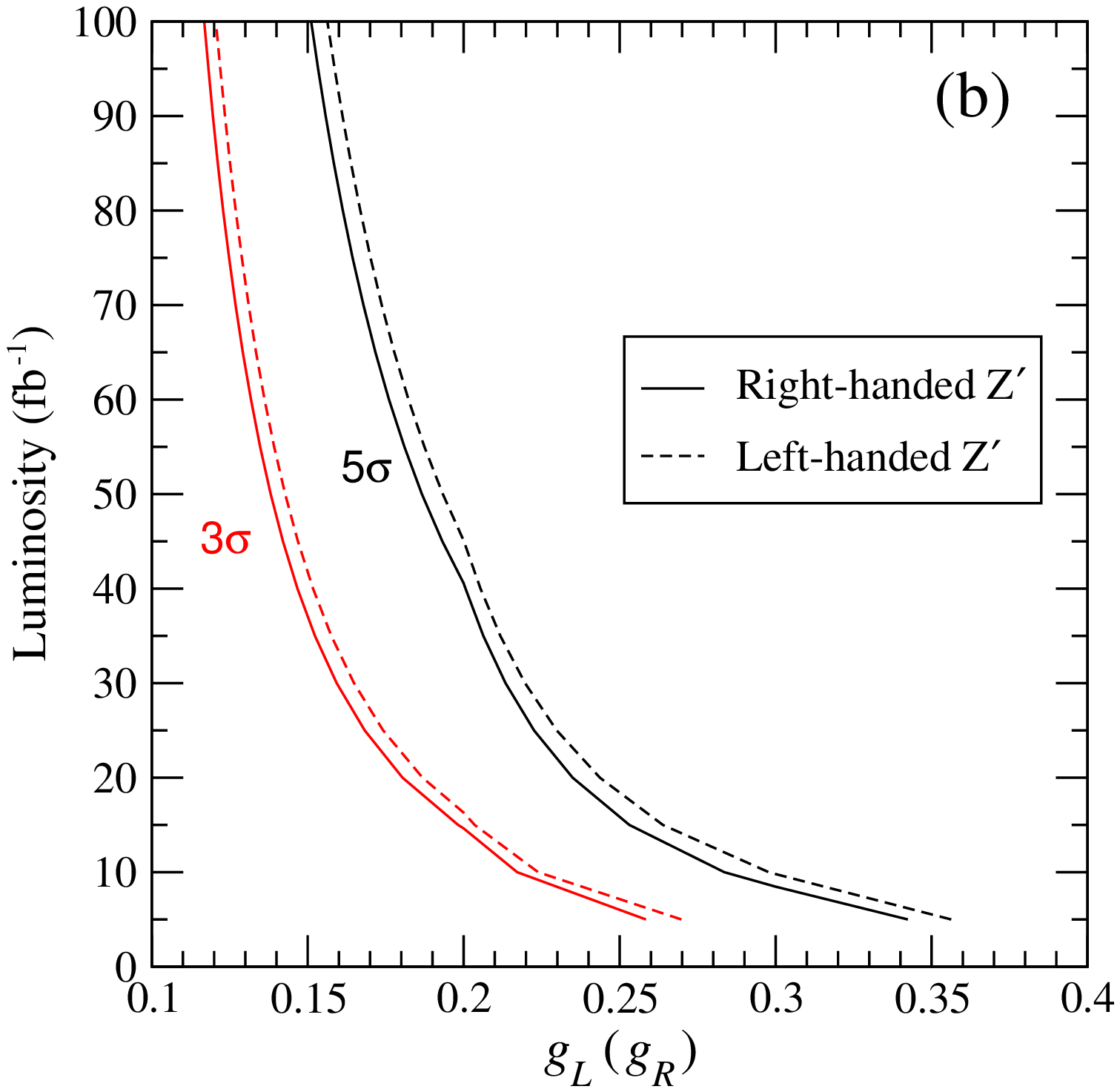}
\caption{(a) Discovery potential for a $Z^\prime$ boson at 14~TeV
as a function of the integrated luminosity for a left-handed $Z^\prime$-$t$-$\bar{t}$ coupling
($g_L=0.4$, $g_R=0$) or a right-handed coupling ($g_L=0$, $g_R=0.4$). 
(b) Luminosities required for 5 standard deviation ($5\sigma$) and 3 standard deviation ($3\sigma$) 
discovery if the $Z^\prime$ mass is 1~TeV, as a function of the coupling strength $g_{L}(g_R)$.  
\label{fig:lhc-potential}}
\end{center}
\end{figure}
Figure~\ref{fig:lhc-potential} shows the statistical significance for discovery as a function of 
accumulated luminosity for $m_{Z^\prime} = 1~{\rm TeV}$ and $m_{Z^\prime} = 1.5~{\rm TeV}$.  
Since the charged lepton from a right-handed top quark decay is boosted to a harder $p_T$, 
more signal events survive after cuts in a right-handed $Z^\prime$ model (black-solid curve) 
than in a left-handed $Z^\prime$ model (red-solid curve).  Therefore, the statistical significance for a right-handed $Z^\prime$ is better when $m_{Z^\prime} = 1~{\rm TeV}$.  However, the situation 
reverses when the $Z^\prime$ becomes heavier because the selection cuts play a role.  For a 
heavy enough $Z^\prime$,  the top quark is highly boosted.  The leptons and jets from its decay are collimated and fail the $\Delta R$ separation cuts, as can be seen in Fig.~\ref{fig:deltaR}.  The 
peak position in the $\Delta R$ distribution for two light jets shifts down below $0.4$ when the mass of a right-handed $Z^\prime$ increases from $1$ TeV to $1.5$ TeV.  As a result, the discovery potential for a very heavy, right-handed $Z^\prime$ is worse than for a left-handed $Z^\prime$, as illustrated in 
the black-dashed and red-dashed curves in Fig.~\ref{fig:lhc-potential} for a 1.5~TeV $Z^\prime$. 
\begin{figure}[t]
\begin{center}
\includegraphics[scale=0.55,clip]{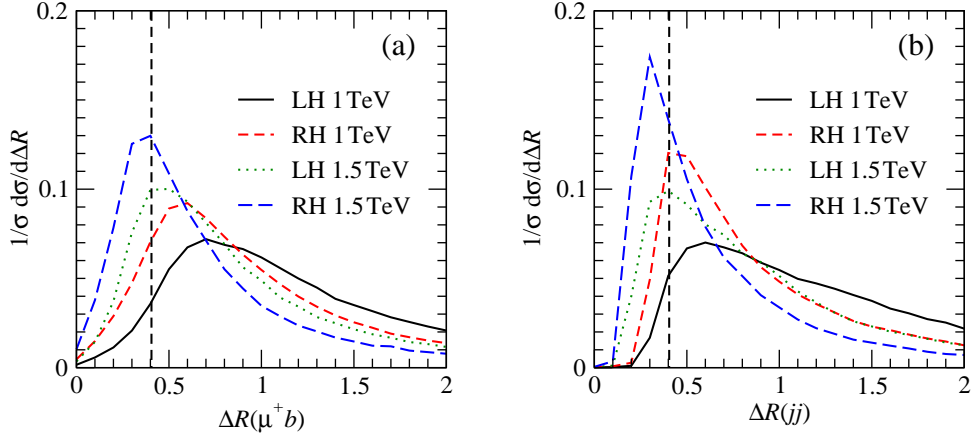}
\caption{(a) $\Delta R$ distributions for the $\mu^+$ and $b$ from $t$ semileptonic decay.  
(b) $\Delta R$ distributions for two light jets from $\bar{t}$ hadronic decay.  The vertical 
dashed line indicates the $\Delta R$ cut applied in our event selection.}
\label{fig:deltaR}
\end{center}
\end{figure}

\subsection{$W^\prime$ discovery potential}

The search for a $W^\prime$ in the $tb$ mode involves the study of  the $\ell^\pm b\bar{b}$ plus 
missing energy final state, where the missing energy originates from the neutrino in top quark 
decay.   We examine only the $W^{\prime+}$ case since its production cross section is about a 
factor of two larger than $W^{\prime-}$.  We demand that two jets are $b$-tagged.  
The SM background processes include production of 
single top quarks,  $Wb\bar{b}$, $Wjj$, and $t\bar{t}$.   Only the 
single-top background is considered in our study because it is about a factor of $10$ larger 
than the others~\cite{Gopalakrishna:2010xm}.
There are three single-$t$ backgrounds: $q\bar{q}^\prime \to t \bar{b}$ (named $t\bar{b}$), 
$q b \to q^\prime t$ (named $bq$) and $q g \to q^\prime t \bar{b}$ (called $Wg$-fusion). 
The $bq$ channel has only one $b$-jet in the final state, but  it is possible that the 
other light-favor jet is misidentified as a $b$-jet as well.
The channel has a large cross section ($\sim 150~{\rm pb}$~\cite{Schwienhorst:2010je}). 
We also apply a mistagging rate for charm-quarks $\epsilon_{c\to b}=10\%$ for 
$p_{T}(c)>50\,{\rm GeV}$.   
The mistag rate for a light jet is $\epsilon_{u,d,s,g\to b}= 0.67\%$ for $p_{T}(j)<100\,{\rm GeV}$
and $2\% $ for $p_{T}(j)>250\,{\rm GeV}$.
For $100\,{\rm GeV}<p_{T}\left(j\right)<250\,{\rm GeV}$, we linearly interpolate the fake rates given above. 

\begin{table}
\caption{Cross sections (in fb) for the signal process $pp\to W^\prime \to t\bar{b} \to \mu^+ \met b \bar{b}$ and the SM backgrounds at 14~TeV.   Two $b$-jets are tagged.  ``With cuts" refers to cross sections after all of the kinematic cuts, b-tagging, and reconstruction. The value $0.4$ is used for the universal coupling of the $W^\prime$ to the SM quarks.  The mass window cuts are 
$\Delta M = 150~(200)$ GeV for a $1~(1.5)$~TeV $W^\prime$, respectively.}
\begin{tabular}{c|ccc|ccc|ccc}
\hline
   &  & $W_R^\prime$&  & &$W_L^\prime$& & &Background & \tabularnewline
\hline  
$M_{W^\prime}$& No cut & With cuts & $\Delta M$ & No cut & With cuts & $\Delta M$ & No cut             & With cuts & $\Delta M$ \tabularnewline
\hline
1~TeV         & 652.1    & 109.4      & 105.2     & 650.3  &112.9 	   & 108.5      &  $3.04\times 10^4$  &   412.1   &  8.4 \tabularnewline
1.5~TeV      & 131.7    & 23.9        & 22.7       & 129.2  & 26.1      & 24.8       & $3.04\times 10^4 $&   412.1   &  2.0 \tabularnewline
\hline
\end{tabular}
\label{table:xswprime}
\end{table}%

All signal and background events are required to pass the acceptance cuts:
\bea
p_T(\ell) > 20~{\rm GeV},&\quad |\eta(\ell)|<2.5, &\quad \met > 25~{\rm GeV},\nonumber \\
p_T(j)>50~{\rm GeV}, &\quad |\eta(j)|<3.0, &\quad \Delta R(jj)>0.4,\quad \Delta R(j\ell)>0.3~.
\eea
The notation is the same as in the $Z^\prime$ search.   The event reconstruction is done in a manner 
similar to Ref.~\cite{Gopalakrishna:2010xm}.  

A two-fold solution may be obtained for the undetected $z$ component of the momentum of the 
missing neutrino if the $W$-boson on-shell condition is used.  However, unlike $Z^\prime \to t \bar{t}$,  there is not a second top quark in the final state to help in selecting the correct $b$-jet to pair with the charged lepton.   Instead, we use the top quark on-shell condition, $m_{b\ell\nu}^2 = m_t^2$ and loop over the two $b$-jets to find the better paring of the $b$-jet and charged lepton, and the better solution 
for the neutrino momentum.  If no real solution is obtained, we discard the event.   After the neutrino 
momentum is obtained, we can reconstruct the momenta of the top quark and the $W^\prime$.  To further suppress the SM backgrounds, we limit the event set to a mass window around the $W^\prime$ 
peak position.  For a $1$ TeV ($1.5$ TeV) resonance, we adopt  
\be
\left|m_{t\bar{b}}-m_{W^\prime}\right|\leq 150~(200)~{\rm GeV}.
\ee
After the cuts are imposed and the two b-jets are tagged, the SM backgrounds are at the fb 
level while the signal rates are about 100~(20)~fb for $W^\prime$ masses of $1$ ($1.5$) TeV.  
The  signal and background cross sections are shown in Table~\ref{table:xswprime}.   
After imposing the mass window cut and demanding two $b$-tagged jets, we find that 
the $Wg$-fusion channel yields the largest background. 
For a 1~TeV $W^\prime$, the single-$t$ processes give these background rates:
$Wg$-fusion, 6.3~fb; $bq$, 1.3~fb; $t\bar{b}$, 0.8~fb.
For a 1.5~TeV $W^\prime$, the $Wg$-fusion rate is 1.4~pb, $bq$ is 0.4~fb, and $t\bar{b}$ is 0.16~pb.   
The acceptance 
for a left-handed $W^\prime$ is slightly better than for a right-handed one.  Because the top quark is boosted from $W^\prime$ decay, $\Delta R$ between the charged lepton and the $b$ quark from 
the decay of a right-handed top quark is smaller than that from a left-handed top quark,  similar to the situation described above for a $1.5$ TeV $Z^\prime$.

The coupling strength $g_L^{W^\prime}=0.4$ or $g_R^{W^\prime}=0.4$ is used in our analysis to satisfy constraints from the Tevatron dijet search.
Since the background is much smaller than the signal rate, $S/B\simeq {\cal O}(10)$, a $W^{\prime+}$ should be easy to discover. Moreover, good accuracy can be obtained for the top quark polarization measurement, as discussed in the next section. %

%
%
%
\section{Top quark polarization}
\label{sec:pol}

The symmetry breaking patterns mentioned in Sec.~II prefer either a purely left-handed
top quark ($SU(2)_1\times SU(2)_2 \to SU(2)_L$) or a purely right-handed top quark 
($SU(2)_R\times U_Y^\prime \to U(1)_Y$).  We can measure the top quark polarization 
from the $\cos\theta$ distribution of the charged lepton in top quark decay after the top 
quark kinematics are reconstructed in the $W^{\prime+}$  and $Z^{\prime+}$ final states, 
as described in Sec.~III.  

\begin{figure}[t]
\begin{center}
\includegraphics[scale=0.7,clip]{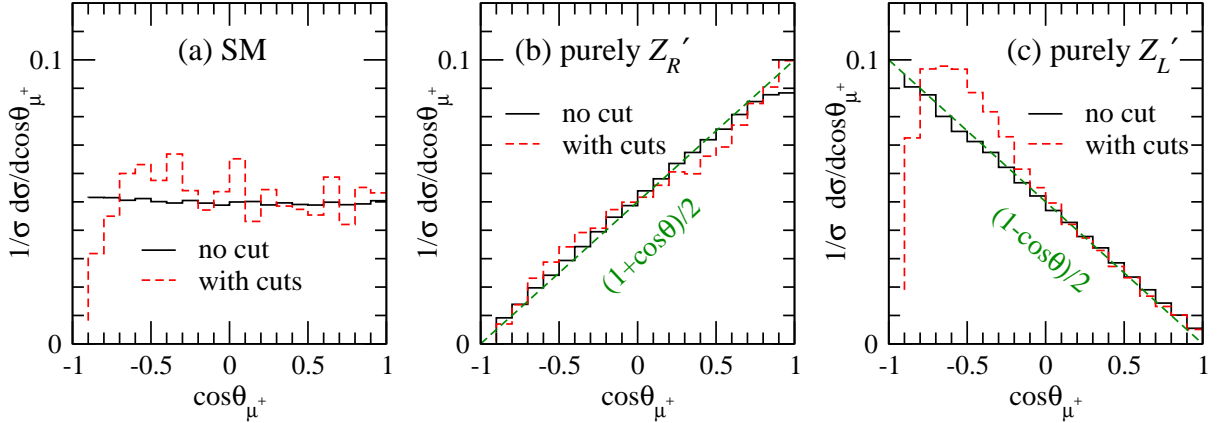}
\caption{Distributions in $\cos\theta_\ell$ of the lepton from the decay of top quarks 
produced in $t\bar{t}$ events before and after cuts: (a) SM, (b) right-handed polarized top 
quarks in $Z^\prime$ decay;  (c) left-handed polarized top quarks in $Z^\prime$ decay.  
The distributions for $W^\prime$ decay are similar to those for $Z^\prime$ decay.
\label{fig:cth_hel_z}}
\end{center}
\end{figure}

Figure~\ref{fig:cth_hel_z} displays the $\cos\theta_\ell$ distributions of the $\mu^+$ in the 
rest frame of the top quark in $t\bar{t}$ events for the SM background and for a 1~TeV 
$Z^\prime$ boson before and after cuts.  A top quark produced at the LHC via QCD 
interactions is unpolarized, as shown by the flat black curve in Fig.~\ref{fig:cth_hel_z} (a).
We see the $1\pm\cos\theta_\ell$ behaviors in Fig.~\ref{fig:cth_hel_z}(b) and (c) for purely 
right- and left-handed polarized top quarks from $Z^\prime$ decay.  After kinematic cuts are 
imposed, the distributions are distorted and drop significantly in the region $\cos\theta_\ell \sim -1$,  affected mainly by the $p_T$ and $\Delta R$ cuts.   However, the main characteristic features remain, i.e. flatness and $1\pm\cos\theta_\ell$.  While not shown here, the $\cos\theta_\ell$ distributions in right- and left-handed $W^\prime$ decay are similar to the $Z^\prime$ case.

We denote the distributions from purely left-handed (right-handed) top quarks $F_L(y)$ 
($F_R(y)$) where $y=\cos\theta_\ell$.  These are the distributions in 
Fig.~\ref{fig:cth_hel_z} (b) and (c) with cuts.  We use these as the basis functions to 
fit the event distributions from our simulations of various models.   We adopt a general 
linear least squares fit in this study to estimate how well the degree of top quark polarization 
can be determined.

An observed angular distribution $O(y)$ after the SM background is subtracted can be expressed as 
\be
O(y) = \epsilon_L~F_L(y) + \epsilon_R~F_R(y),
\ee
where $\epsilon_L$ ($\epsilon_R$) is the fraction of left-handed (right-handed) top quarks. 
The values of $\epsilon_L$ and $\epsilon_R$ are chosen as the best parameters that minimize $\chi^2$, defined as
\be
\chi^2=\sum_{i=1}^N \left[\frac{O(y_i)-\epsilon_L F_L(y_i)-\epsilon_R F_R(y_i)}{\sigma_i}\right]^2,
\label{eq:chi2}
\ee 
where $N$ is the number of bins, and $\sigma_i = \sqrt{O(y_i)}$ is the statistical error 
(standard deviation) of the $i$th data point.
The minimum of Eq.~\ref{eq:chi2} occurs where the derivative of $\chi^2$ with respect to 
both $\epsilon_L$ and $\epsilon_R$ vanishes, yielding the normal equations of a least-squares problem:
\be
0 = \sum_{i=1}^{N} \frac{1}{\sigma_i^2} \left[O(y_i)-\epsilon_L F_L(y_i)-\epsilon_R F_R(y_i)\right] F_l(y_i),
\quad {\rm where}~~l=L(R).
\ee
Interchanging the order of summations, one can write the above equations as matrix equations, 
\be
\alpha_{LL} \epsilon_L + \alpha_{LR} \epsilon_R = \beta_L, \qquad \alpha_{RL} \epsilon_L + \alpha_{RR} \epsilon_R = \beta_R,
\label{normaleq}
\ee
where
\be
\alpha_{lm}=\sum_{i=1}^{N}\frac{F_l(y_i)F_m(y_i)}{\sigma_i^2},\qquad
\beta_l = \sum_{i=1}^N \frac{O(y_i)F_l(y_i)}{\sigma_i^2}.
\ee
The coefficients $\epsilon_L$ and $\epsilon_R$ can be obtained from Eq.~\ref{normaleq} as
\be
\epsilon_l=\sum_{m=L}^{R} [\alpha_{lm}]^{-1}\beta_{m}=\sum_{m=L}^{R} C_{lm}
\left[\sum_{i=1}^{N}\frac{O(y_i)F_m(y_i)}{\sigma_i^2}\right],\quad l=L,R~.
\ee
The inverse matrix $C_{lm}\equiv [\alpha_{lm}]^{-1}$ is closely related to the standard uncertainties of the estimated coefficients $\epsilon_L$ and $\epsilon_R$.   Assuming the data points are independent, 
consideration of propagation of errors shows that the variance $\sigma_f^2$ in the value of $\epsilon_l$ is
\be
\sigma^2(\epsilon_l)= \sum_{i=1}^{N} \sigma_i^2 \left(\frac{\partial \epsilon_l}{\partial O(y_i)}\right)^2=C_{ll},
\ee
i.e. the diagonal elements of $[C]$ are the variances (squared uncertainties) of the fitted coefficients. 

We  illustrate this method with two toy models in which the $W^\prime tb$ and $Z^\prime t t$  couplings  are $\epsilon_L=30\%$ and $\epsilon_L =  70\%$, where the fraction of left-handed top quarks is closely related to the $W^\prime$-$t$-$\bar{b}$ and
$Z^\prime$-$t$-$\bar{t}$ couplings as
\be
\epsilon_L \equiv \frac{\sigma(t_L)}{\sigma(t_L)+\sigma(t_R)} \approx \frac{g_L^2}{g_L^2 + g_R^2}.
\ee
For simplicity, we assume identical $\epsilon_L$ for $W^\prime$ and $Z^\prime$.  (Note that 
$\epsilon_L  \approx 1 - \epsilon_R$.)   
The coupling strength to the top quark ($g_V=\sqrt{g_L^2+g_R^2}$) and the masses of 
the $Z^\prime$ and $W^\prime$ are taken to be the same: $g_V^{Z^\prime}=g_V^{W^\prime} = 0.4$; 
$m_{Z^\prime} = m_{W^\prime} = 1~{\rm TeV}$.     
We also adopt $5\%$ uncertainties in each bin to take into account the imperfect predictions of the template distributions $F_L(y)$ and $F_R(y)$.~\footnote{The $5\%$ variation may be too optimistic, 
but the value to be adopted will not be obvious until a more precise next-to-leading order calculation 
is done.}
After including SM backgrounds, we generate ten 
bins of data in the $\cos\theta_\ell$ distribution.  The first  bin, $\cos\theta_\ell\sim -1$, is not used in 
the fits because of the significant drop-off associated with cuts.  

\begin{figure}[t]
\begin{center}
\includegraphics[scale=0.55,clip]{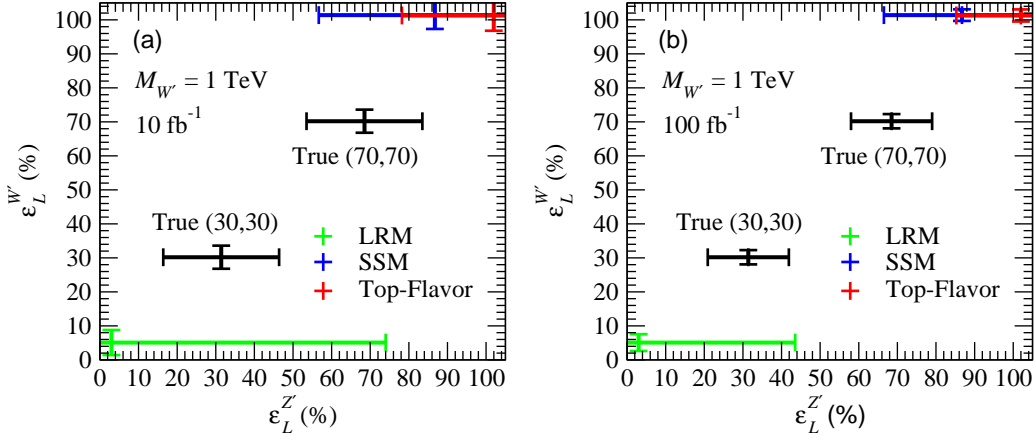}
\caption{Top quark polarizations determined from $\chi^2$ fits in 
$pp\to W^{\prime+} \to t\bar{b}$ and $pp\to Z^\prime \to t\bar{t}$ for two assumed 
$t$-polarizations  $\epsilon_L=0.3$, 0.7.  The results for benchmark models, the 
left-right symmetric model (LRM), the sequential standard model (SSM where couplings 
are the same as the SM $W$ and $Z$ bosons), and the top-flavor model, are also shown.  
The uncertainties are the quadratic sum of the uncertainties from statistics and theory and 
are shown as black and red bands for integrated luminosities of $10~{\rm fb}^{-1}$ and 
$100~{\rm fb}^{-1}$, respectively.
\label{fig:dop}}
\end{center}
\end{figure}

The results are shown in Fig.~\ref{fig:dop} with the legends ``True~(30,30)" and ``True~(70,70)".   For an assumed $\epsilon_L = 30\%~(70\%)$ used in  the event generation, the measured polarization ($\epsilon_L^{mea}$) is found to be $\epsilon_L^{mea} = 31.4\% \pm 15\%~(68.5\%\pm15\%)$ for a $Z^\prime$, and 
$\epsilon_L^{mea} = 30.2\% \pm 3.4\%~(70.2\%\pm 3.4\%)$ for a $W^\prime$ for an integrated luminosity of  $10~{\rm fb}^{-1}$.   The uncertainties are reduced to $10.5\% $ and $2.1\%$ for a $Z^\prime$ and a $W^\prime$, respectively, if the integrated luminosity is raised to $100~{\rm fb}^{-1}$, while the central values remain the same.   As shown in the figure, $\epsilon_L$ can be measured precisely in the 
$W^\prime \to t\bar{b}$ mode, owing to the small SM backgrounds, while the measurement in the 
$Z^\prime\to t\bar{t}$ channel is less accurate because of the large $t\bar{t}$ background. 

The statistical uncertainty of each bin in the $\cos\theta_\ell$ distribution scales as 
$\sqrt{N_i}=\sqrt{\sigma_i \mathcal{L}}$ where $\sigma_i$ is the 
differential cross section of the $i$-th bin.  An increase of the luminosity by a factor of 
$k$ reduces the uncertainties by a factor of $\sqrt{k}$.

Applying our method to the three benchmark NP models described in Sec.~II, we obtain the 
results shown in Fig.~\ref{fig:dop} for $m_{W^\prime} = 1~{\rm TeV}$.  In the SSM and top-flavor
models $m_{Z^\prime} = m_{W^\prime}$, while in the LRM $m_{Z^\prime}\simeq 1.2~m_{W^\prime}$ for 
$\alpha_{LR}=1.6$.  
All the fitted central values are very close to the true values in each model. 

Because SM backgrounds are relatively small, the polarization of the top quark in $W^\prime$ 
decay can be measured precisely.  The uncertainty is $\lesssim 5\%$ in $\epsilon_L^{W^\prime}$ 
for $10~{\rm fb}^{-1}$ of integrated luminosity.  It decreases to $\lesssim 3\%$ for $100~{\rm fb}^{-1}$ of integrated luminosity.   The polarization is measured less accurately in the $t\bar{t}$ channel, however, 
because the SM $t\bar{t}$ background is large.   For a $Z^\prime$ in the SSM and top-flavor models, the uncertainties are  about $30\%$ and $24\%$ 
with $10~{\rm fb}^{-1}$ of integrated luminosity, respectively.  
They are reduced to about $20\%$ and $17\%$ when the integrated luminosity is 
increased to $100~{\rm fb}^{-1}$. 
For a $Z^\prime$ in the LRM, the uncertainty is large,  
$\sim 70\%$ for $10~{\rm fb}^{-1}$ and $\sim 40\%$  for $100~{\rm fb}^{-1}$ integrated luminosities.  
The larger uncertainties arise because $Z^\prime$ mass is heavier ($1.2$ TeV) and the coupling 
of the $Z^\prime$ to $t\bar{t}$ is smaller than in the SSM and top-flavor models, as noted in 
Table~\ref{table:coup}.  The statistics are too low to obtain a good fit.  

The fitted uncertainties are generally larger for a left-handed $Z^\prime_L$ compared with a right-handed  $Z^\prime_R$ if the the masses and couplings to the top quark are the same, because the charged lepton from the decay of a left-handed top quark is softer than from a right-handed top quark. Therefore, the statistics are usually greater for a $Z^\prime_R$.  However, the situation switches when the $Z^\prime$ is heavier, because the separation $\Delta R$ computed from the charged lepton and the $b$ from top quark decay will more easily fail the kinematic cuts, as shown in Sec.~\ref{sec:lhc_z}. 

\section{Discussion and Summary}

Extra gauge bosons may be among the new states produced at the LHC.   Depending on how the 
gauge group is broken in the BSM scenario and how the SM fermions are charged, the couplings of 
the heavy gauge bosons $W^\prime$ and $Z^\prime$ to the SM fermions could have 
the same or different handedness.   
While  it may be more straightforward to discover the $W^\prime/Z^\prime$ through their dijet or leptonic decays, we emphasize that decays into top quark final states allow us to study the nature of the 
coupling between the new gauge bosons and the SM quarks.  These channels are also 
complementary to the dijet and leptonic channels, especially for leptophobic $W^\prime/Z^\prime$ bosons.   

We focus on final states in which the $W^{\prime+}$ boson decays into a top quark and a bottom quark and 
the $Z^\prime$ boson decays into a top-antitop quark pair.  The collider signatures we examine are 
{$\ell^+ \met b\bar{b}$} for the $W^{\prime +}$ and the semileptonic channel $\ell^\pm \met j j b \bar{b}$ for the $Z^\prime$.   The missing energy $\met$ is carried off by a neutrino in the top quark decay.  
After an event simulation of the signals and backgrounds, and imposing a set of kinematic cuts,  we reconstruct the momentum of the missing neutrino using on-shell conditions for the $W$ boson 
and the top quark.  Adopting a coupling strength $g_V=\sqrt{g_L^2+g_R^2}=0.4$ 
($g_L$ and $g_R$ are the left-handed and right-handed $Z^\prime t\bar{t}$ couplings) 
consistent with data from the Tevatron 
dijet search, we find that a $1$ TeV $Z^\prime$ resonance could be seen above the SM 
$t\bar{t}$ background at $14$ TeV with a  statistical significance of more than $5 \sigma$ for 
$10~{\rm fb}^{-1}$ of integrated luminosity.   The $W^\prime$ signal can be much larger than the 
SM background if the coupling strength of the $W^\prime$ to the SM quarks is not too small ( $\sim 0.4$). 

Even in the presence of SM backgrounds and despite distortions associated with cuts, we  
show that the left-right handedness of the top quark can be measured from the angular distribution 
in $\cos\theta_{\ell}$ of the charged lepton in the top quark rest frame.   This observable is most interesting because it reflects the coupling structure of the $W^\prime/Z^\prime$ to the top quark.  
When performing $\chi^2$ fits to our simulated $\cos\theta_{\ell}$ distributions, we allow $5\%$ fluctuations in each bin of $\cos\theta_{\ell}$. In the $W^\prime$ case, the SM backgrounds are 
negligible and our fits result in uncertainties of $\lesssim 5\%$ for the fraction of left-handed  
coupling for $10~{\rm fb}^{-1}$  of accumulated luminosity.  The $Z^\prime$ situation is less good because the statistics are smaller and the SM backgrounds are larger.  The uncertainty for the 
$Z^\prime$  decreases from $\sim 15\%$ to $\sim10\%$ when the integrated luminosity is increased  
from $10~{\rm fb}^{-1}$ to  $100~{\rm fb}^{-1}$.  

We apply our approach to three benchmark models, the SSM, LRM, and top-flavor models.  
The central values from our fits are very close to the true values in these models, with a deviation 
about $2\%\sim 5\%$, and they are insensitive to luminosity.  The uncertainties, again, are small for $W^\prime$ and larger for $Z^\prime$, depending mainly on statistics. 
Owing to the larger uncertainty in 
the $Z^\prime$ case and the similarity in the handedness of couplings (true values $\epsilon_L\simeq 84\%$ for SSM and $\epsilon_L = 100\%$ for top-flavor model), the SSM and top-flavor models can be separated 
only marginally with $100~{\rm fb}^{-1}$ of integrated luminosity when the $Z^\prime$  and 
$W^\prime$ results are combined, as shown in Fig.~\ref{fig:dop}. However, the situation is better if the $Z^\prime$ is leptophobic and the coupling strength to SM quarks is as large as $0.4$.  This is the case we show first using toy models with the couplings $\epsilon_L = 30\%$ and $70\%$ in Fig. 7.  
Since the uncertainty can be reduced to about $10\%$, one could distinguish models with  handedness of couplings to SM quarks differing by about $10\%$. 

We remark that the coupling strength of a $W^\prime/Z^\prime$ to the top quark cannot be 
determined from our study since our signal cross sections also depend on the couplings of 
the $W^\prime/Z^\prime$ to light quarks.  Finally, our approach relies on reliable calculations 
for the backgrounds as well as the signals, which are crucial for normalizations and the kinematic distributions.   Next-to-leading order (NLO) QCD contributions enhance the cross section 
significantly~\cite{Sullivan:2002jt}, possibly improving our results.   The higher order QCD contributions  to the SM backgrounds are also not negligible.   A study that consistently includes higher order contributions to both the backgrounds and the signals is needed, and we leave it for future work.

\begin{acknowledgments}
We thank Jiang-Hao Yu for fruitful discussions.
The work by E.L.B. and Q.H.C. is supported in part by the U.S. DOE
under Grants No.~DE-AC02-06CH11357. Q.H.C. is also
supported in part by the Argonne National Laboratory and University
of Chicago Joint Theory Institute Grant 03921-07-137. C.R.C. is supported
by World Premier International Initiative, MEXT, Japan. H.Z. is supported
in part by the National Natural Science Foundation of China under
Grants 10975004 and the China Scholarship Council File No. 2009601282. C.R.C. thanks NCTS and Institute
of Physics, Academia Sinica in Taiwan for the hospitality during the final stages
of this work.
\end{acknowledgments}

\end{document}